# Harris criterion on hierarchical lattices: Rigorous inequalities and counterexamples in Ising systems

Avishay Efrat

School of Physics and Astronomy
Tel Aviv University,
Ramat Aviv, Tel Aviv 69978, Israel.

Random bond Ising systems on a general hierarchical lattice are considered. The inequality between the specific heat exponent of the pure system, $\alpha_p$, and the crossover exponent $\phi$, $\alpha_p \leq \phi$, gives rise to the possibility of a negative $\alpha_p$ along with a positive $\phi$, leading to random criticality in disagreement with the Harris criterion. An explicit example where this really happens for an Ising system is presented and discussed. In addition to that, it is shown that in the presence of full long-range correlations the crossover exponent is larger than in the uncorrelated case.

One of the most famous and important results in the study of systems affected by quenched disorder is the Harris criterion [1]. The widely accepted form of the criterion, is that in ferromagnetic systems with random interactions the randomness is irrelevant if $\alpha_p$, the specific heat exponent of the corresponding pure system, is negative, while for systems with positive $\alpha_p$ the random system exhibits different critical behavior. Specifically, $\phi = \alpha_p$, where $\phi$ is the crossover exponent from pure to random criticality. It is surprising, however, that in spite of the long time since its proposal and quite a number of alternative derivations [2-10], a rigorous proof of the Harris criterion is still lacking. In fact, the only rigorous results in this field [11,12] yield information about the exponents of the random system ($\alpha_r < 0$ or $\nu_r > 2/d$, $d$ being the dimension of the system) without relating those to the exponents of the pure system. Some years ago, a counterexample to the Harris criterion was presented for Potts models on a certain hierarchical lattice (HL) [13]. It was shown that, for a $q$-state Potts model, it is possible to find a window of rather large $q$'s in which $\alpha_p$ is negative but the disorder is relevant. It is unclear whether the Harris criterion is also violated in the Ising model, because in contrast with high $q$ Potts models [8] it is quite difficult to obtain negative $\alpha_p$'s in Ising systems.

The above situation motivated the present work, which yields some rigorous results regarding the Harris criterion for the Ising model. Because on regular lattices any attempt to obtain some rigorous results via a renormalization group (RG) procedure at the pure fixed point is hampered by the generation of correlations, I chose to work on general HLs. This produces exact results because correlations are not generated in such systems under RG (note that some derivations of the Harris criterion [8-10] assume the same in one way or another). The results I obtain here are the following:

(1) I define two random interactions to be fully correlated if both are identical in each representation of the randomness. I define the full correlations as long ranged if they survive the renormalization procedure for any number of steps. In the presence of full long-range correlations, the crossover exponent, to be denoted as $\phi_{corr}$, is larger than $\phi$, the crossover exponent of the uncorrelated system.

(2) Explicit examples are constructed where $\alpha_p$ is negative but still the randomness is relevant. This is true for uncorrelated and the correlated systems.

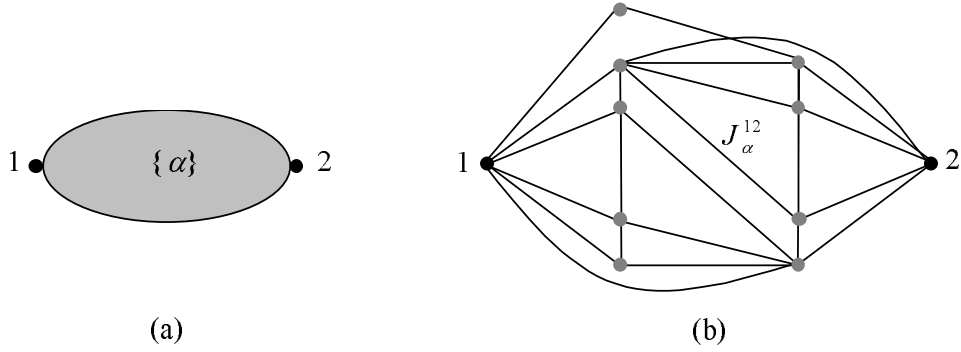

**Fig. 1.** A general HL is described schematically. In (a), the shaded area consists of a set of lattice points, where some of the pairs are joined by bonds $\alpha, \beta, \ldots$. In (b), a typical shaded area is represented. The full lines are bonds to be iterated in constructing the lattice.

Consider a general HL described schematically in Fig. 1. The shaded area shown in (a) consists of a set of lattice points where some of the pairs are joined. In (b), a typical shaded area is represented. The lines are bonds to be iterated in constructing the lattice. All bonds carry a coupling $J_\alpha^{12}$, governed by a distribution $P(J_\alpha)$ that is identical for all bonds. The renormalized coupling is given by

$$\tilde{J}_{12} = f\{J_\alpha^{12}\}, \tag{1}$$

where $f$ depends only on couplings associated with the pair of sites (1,2) (the shaded area, Fig. 1). This implies that $\tilde{J}_{ij}$ and $\tilde{J}_{lm}$ are not correlated if the pairs $(i,j)$ and $(l,m)$ are not identical. The renormalized distribution $\tilde{P}(\tilde{J})$ is given by

$$\tilde{P}(\tilde{J}_{12}) = \int \prod_\alpha dJ_\alpha^{12} P(J_\alpha^{12}) \delta[\tilde{J}_{12} - f\{J_\beta^{12}\}] \tag{2}$$

and may serve to derive an infinite set of equations for the renormalized moments. Let us denote

$$\Gamma_i = \langle (\delta J_\alpha)^i \rangle, \tag{3}$$

where $\delta J_\alpha$ denotes the departure from the pure fixed point $J_\alpha^{12} = J^*$ for all $\alpha$. The recursion equations for the moments read

$$\tilde{\Gamma}_i = G_i[\Gamma_1, \Gamma_2, \ldots]. \tag{4}$$

The pure ferromagnetic fixed point is assumed at $J^* > 0$, so that

$$\Gamma_i^* = 0. \tag{5}$$

In a recent paper [14], the following results were proved by considering the matrix $\left(\partial \tilde{\Gamma}_i / \partial \Gamma_j\right)(0,0,...)$.

(a) The eigenvalues of the matrix, $\lambda_i$, are given by

$$\lambda_i = \sum_{\alpha=1}^{n} (f_\alpha)^i \qquad \text{where} \qquad f_\alpha \equiv \frac{\partial f}{\partial J_\alpha}(J^*,...,J^*). \qquad (6)$$

The sum is over all $n$ bonds $\alpha$ of the rescaling volume associated with the pair (1,2) (the superscript "1,2" is omitted hereafter) and the partial derivative is taken at the point where all those couplings equal $J^*$.

(b) All the eigenvalues are positive.

(c) $\lambda_{i+1} < \lambda_i$. $\qquad (7)$

As a consequence of (c), the leading crossover exponent is

$$\phi_2 = \frac{\ln \lambda_2}{\ln \lambda_1} \qquad (8)$$

and

$$\phi_2 \geq \alpha_p, \qquad (9)$$

where $\alpha_p$, the specific heat exponent of the pure system, may be easily obtained by consideration of the free energy per bond,

$$2 - \alpha_p = \frac{\ln n}{\ln \lambda_1}. \qquad (10)$$

The equality sign in (9) holds only if all the bonds in the shaded area of Fig. 1(a) are equivalent.

We turn now to the case where full long-range correlations are present in the system. In the following, I define full long-range correlations for a general HL. Consider a HL, generated using some specific generator such as the one presented in Fig. 2(a), where as before, the different $J_\alpha$'s ($\alpha = 1,...,5$ here) represent the independent random interactions attached to the different bonds. In Fig. 2(b), some of these bonds are taken to be fully correlated, namely, bonds $\alpha$ and $\beta$ are fully correlated if, say, $J_\beta = J_\alpha$ so that indeed there is only one independent random variable, say $J_\alpha$. In order for this property to survive the renormalization procedure, the bonds have to be distributed in such a way that this property is true on all scales [as demonstrated in Fig. 2(c)]. In this way, correlations are nonlocal at all scales and thus long ranged.

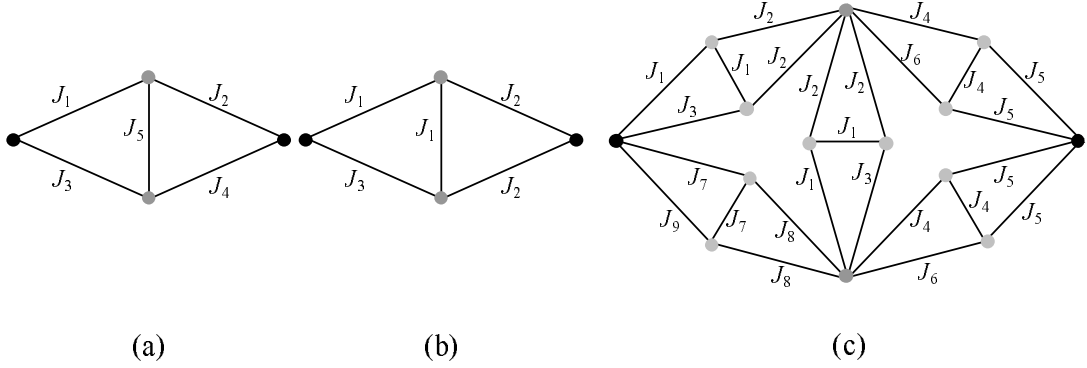

(a)　　　　　　　　(b)　　　　　　　　(c)

**Fig. 2.** In (a), an example of a HL, generated using some specific generator, is given. The different $J_\alpha$'s ($\alpha = 1,\ldots,5$ here) represent the different random interactions attached to the bonds. In (b), some of the bonds associated with that generator are taken to be fully correlated, that is, the interactions $J_\alpha$ attached to these bonds are taken to be identical. This must hold at all scales in order for the correlations to be nonlocal and survive the renormalization procedure. In (c), we zoom in on one renormalization step of a possible configuration that generates such correlations. The bonds are distributed in such a way that the initial setup of correlated bonds defined on the generator shown in (b) remains true on all scales. In this way, correlations are nonlocal on all scales and thus long ranged.

Let us assume then that the $n$ bonds inside a rescaling volume are divided into $m$ subsets $\Omega_t$, with $t = 1,\ldots,m$ and $1 \leq m \leq n$, each containing $n_t$ fully correlated bonds, so that $\sum_{t=1}^{m} n_t = n$. We can write

$$f\{J_\alpha\} = g\{J_t\}, \tag{11}$$

because all the couplings in $\Omega_t$ equal $J_t$. Finally, we use Eq. (6) for the $\lambda_i$'s to write

$$\lambda_i^{\text{corr}} = \sum_{t=1}^{m} (g_t)^i = \sum_{t=1}^{m} \left( \sum_{\alpha \in \Omega_t} f_\alpha \right)^i \geq \sum_{\alpha=1}^{n} (f_\alpha)^i = \lambda_i \text{ (uncorrelated)}. \tag{12}$$

This is true since $f_\alpha \geq 0$ for all $\alpha$'s [14]. The inequality $\phi \leq \phi_{\text{corr}}$ thus immediately follows and the equality sign holds only if no correlations are present. A special case of the above inequality was obtained a long time ago by Andelman and Aharony [10]. Using the Migdal-Kadanoff (MK) renormalization scheme, they considered $d$-dimensional cubic systems with quenched bond disorder that are correlated along $d_1$ dimensions and showed that the crossover exponent and thus the Harris criterion are modified, respectively, to $\phi_{\text{corr}} = \alpha_p + d_1 \nu_p$ and $2 - (d - d_1)\nu_p < 0$. Since the MK transformation performed on a regular lattice corresponds to an exact RG transformation performed on a diamond HL [15,16], for which all bonds are equivalent and thus $\alpha_p = \phi$, their result reads $\phi_{corr} = \phi + d_1 \nu \geq \phi$.

In the following I present an Ising system, for which $\alpha_p$ is negative while $\phi_2$ is positive. Consider the HL generated using the generator presented in Fig 3. It is made of $p$ parallel branches, each of which contains a single site, connected on one side with a single bond and on its other side with $q$ parallel bonds. The rescaling volume is then $n = p(q+1)$ and, denoting $\beta J_\alpha \equiv K_\alpha$, the renormalized coupling is given by

$$\widetilde{K} = f\{K_\alpha, K_{\alpha\beta}\} = \frac{1}{2}\sum_{\alpha=1}^{p} \ln\left[\frac{\cosh\left(K_\alpha + \sum_{\beta=1}^{q} K_{\alpha\beta}\right)}{\cosh\left(K_\alpha - \sum_{\beta=1}^{q} K_{\alpha\beta}\right)}\right], \quad (13)$$

where $K^*$ is determined by

$$K^* = \frac{1}{2}\sum_{\alpha=1}^{p} \ln\left\{\frac{\cosh[(q+1)K^*]}{\cosh[(q-1)K^*]}\right\}. \quad (14)$$

Taking the first partial derivatives of Eq. (13) with respect to $K_\alpha$ and $K_{\alpha\beta}$, and using Eq. (6), the following expression is obtained for the eigenvalues;

$$\lambda_i = \frac{p}{2^i}\left\{\left(\tanh[(q+1)K^*] - \tanh[(q-1)K^*]\right)^i + q\left(\tanh[(q+1)K^*] + \tanh[(q-1)K^*]\right)^i\right\}. \quad (15)$$

$\alpha_p$ and $\phi_2$ are next calculated according to Eqs. (8) and (10). Numerical results for $K^*$, $\alpha_p$ and $\phi_2$ are given in Table 1 for different values of $p$ and $q$. Note the cells emphasized in gray, where the Harris criterion mistakenly indicates pure criticality.

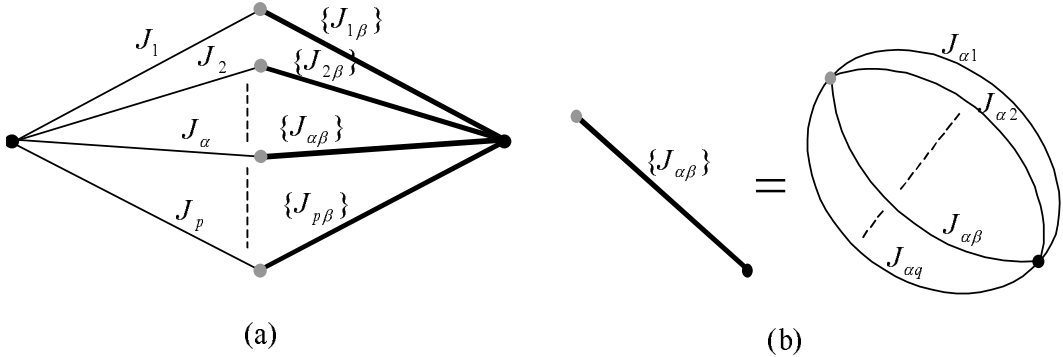

(a)  (b)

Fig. 3. A HL generator is shown that is capable of demonstrating examples for the Ising model for which the Harris criterion fails to predict random criticality. It is made of $p$ parallel branches, each of which contains a single site, connected on one side with a single bond and on its other side with $q$ parallel bonds. In (a), the $p$ parallel branches are shown. Each of the bold bonds represents a set of $q$ parallel bonds as explicitly drawn in (b).

I also considered an example for the case where some of the bonds are fully correlated. For simplicity I have chosen the case in which the $q$ parallel bonds of each of the $p$ branches are fully correlated. Therefore, in each branch, we may replace the $q$ bonds of Fig. 3(b), each carrying a coupling $J_{\alpha\beta}$, with an effective coupling $qJ'_\alpha$, so that now

$$\widetilde{K}_{\text{corr}} = f_{\text{corr}}\{K_\alpha, K'_\alpha\} = \frac{1}{2}\sum_{\alpha=1}^{p} \ln\left[\frac{\cosh(K_\alpha + qK'_\alpha)}{\cosh(K_\alpha - qK'_\alpha)}\right] \qquad (16)$$

and

$$\lambda_i^{\text{corr}} = \frac{p}{2^i}\left\{\left(\tanh[(q+1)K^*] - \tanh[(q-1)K^*]\right)^i + q^i\left(\tanh[(q+1)K^*] + \tanh[(q-1)K^*]\right)^i\right\}. \qquad (17)$$

While $K^*$, $\lambda_1$, and therefore $\alpha_p$, are left unchanged, $\lambda_2^{\text{corr}}$ and therefore $\phi_{\text{corr}}$ are now increased. Numerical results for $\phi_{\text{corr}}$ are also given in Table 1.

| $p$ | 2 | | | 3 | | | | 4 | | | | |
|---|---|---|---|---|---|---|---|---|---|---|---|---|
| $q$ | 1 | 2 | 3 | 1 | 2 | 3 | 4 | 1 | 2 | 3 | 4 | 5 |
| $K^*$ | 0.609 | 0.281 | 0.185 | .361 | .175 | .116 | .087 | .261 | .128 | .085 | .064 | .051 |
| $\alpha_p$ | -.676 | -.111 | .224 | -.902 | -.478 | -.090 | .117 | -1.19 | -.762 | -.319 | -.077 | .069 |
| $\phi$ | -.676 | .012 | .339 | -.902 | -.379 | .013 | .203 | -1.19 | -.670 | -.220 | .006 | .138 |
| $\phi_{\text{corr}}$ | -.676 | .778 | 1.26 | -.902 | .345 | .919 | 1.16 | -1.19 | .040 | .680 | .961 | 1.11 |

**Table 1.** Numerical results for $K^*$, $\alpha_p$, $\phi$, and $\phi_{\text{corr}}$ for different values of $p$ and $q$ with respect to the HL shown in Fig. 3. Note the cells emphasized in gray, where $\alpha_p < 0 < \phi$ so that the Harris criterion mistakenly indicates pure criticality. Also note that always $\alpha_p \leq \phi \leq \phi_{\text{corr}}$.

To conclude, I wish to discuss the possible relevance of my results to regular lattices. One should note that the two properties discussed here, that of bonds inside a rescaling volume not being all equivalent and that of full long-range correlations being present, each by itself makes $\phi_2 > \alpha_p$ possible. This may suggest, at first sight, that starting on a regular lattice with a set of nonequivalent interactions, say, nearest neighbors and next-nearest neighbors, may violate the Harris criterion. On HLs we see, however, that it is also important that these properties are invariant under the RG. On regular lattices, though, we expect that the detailed structure of nonequivalent bonds will disappear under RG and therefore the Harris criterion is expected to hold. Cases where on a regular lattice we may find negative $\alpha_p$ but still randomness is relevant must thus be such that the nonequivalence of bonds does not disappear under RG. This can happen in principle when the interactions are algebraically long ranged or when the correlations of short range interactions are long ranged.

**Acknowledgments:**

I wish to thank professor Moshe Schwartz for many helpful discussions.